\documentclass[conference]{IEEEtran}
\IEEEoverridecommandlockouts
\usepackage{cite}
\usepackage{amsmath,amssymb,amsfonts}
\usepackage{algorithm}
\usepackage{algorithmic}
\usepackage{graphicx}
\usepackage{textcomp}
\usepackage{xcolor}
\usepackage{stfloats}
\usepackage{graphicx}
\usepackage{color}
\usepackage{multirow}
\usepackage{array}

\newcommand{\tabincell}[2]{\begin{tabular}{@{}#1@{}}#2\end{tabular}}
\def\BibTeX{{\rm B\kern-.05em{\sc i\kern-.025em b}\kern-.08em
    T\kern-.1667em\lower.7ex\hbox{E}\kern-.125emX}}
\begin{document}

\title{SPARC-LDPC Coding for MIMO Massive Unsourced Random Access\\}
\author{\IEEEauthorblockN{Tianya Li, Yongpeng Wu, Mengfan Zheng, Dongming Wang, and Wenjun Zhang}
\thanks{{ T. Li, Y. Wu (corresponding author) and W. Zhang are with the Department of
	Electronic Engineering at Shanghai Jiao Tong University, Shanghai, China.
	Emails: \{tianya, yongpeng.wu, zhangwenjun\}@sjtu.edu.cn.

M. Zheng is with the Department of Electrical and Electronic Engineering at
Imperial College London, United Kingdom. Email: m.zheng@imperial.ac.uk.

D. Wang is with the National Mobile Communication Research
Laboratory, Southeast University, Nanjing 210096, China, and also with
Purple Mountain Laboratories, Nanjing 211111, China. Email: wangdm@seu.edu.cn.

The work of Y. Wu is supported in part by the National Key R\&D
Program of China under Grant 2018YFB1801102,  JiangXi Key R\&D Program
under Grant 20181ACE50028, National Science Foundation
(NSFC) under Grant 61701301, the open research project of State Key Laboratory of Integrated
Services Networks (Xidian University) under Grant ISN20-03, and Shanghai Key Laboratory of Digital Media Processing and Transmission (STCSM18DZ2270700).
The work of W. Zhang is supported by  Shanghai Key Laboratory of Digital Media Processing and Transmission (STCSM18DZ2270700).}}
}

\maketitle

\begin{abstract}
A joint sparse-regression-code (SPARC) and low-density-parity-check (LDPC) coding scheme for multiple-input multiple-output (MIMO) massive unsourced random access (URA) is proposed in this paper. Different from the state-of-the-art covariance based maximum likelihood (CB-ML) detection scheme, we first split users' messages into two parts. The former part is encoded by SPARCs and tasked to recover part of the messages, the corresponding channel coefficients as well as the interleaving patterns by compressed sensing. The latter part is coded by LDPC codes and then interleaved by the interleave-division multiple access (IDMA) scheme. The decoding of the latter part is based on belief propagation (BP) joint with successive interference cancellation (SIC). Numerical results show our scheme outperforms the CB-ML scheme when the number of antennas at the base station is smaller than that of active users. The complexity of our scheme is with the order $\mathcal{O}\left(2^{B_p}ML+\widehat{K}ML\right)$ and lower than the CB-ML scheme. Moreover, our scheme has higher spectral efficiency (nearly $15$ times larger) than CB-ML as we only split messages into two parts.
\end{abstract}


\section{Introduction}
Future wireless cellular networks aim to support massive connectivity scenarios such as Internet-of-Things (IOT) and massive machine-type communications (mMTC), etc. A key feature of these scenarios is that the cellular base station (BS) needs to serve a massive number of users or devices\cite{wu2020massive}. However, the pilot-based scheme investigated in \cite{8323218} will cause a waste of pilot resources because of the massive number of users and users' sporadic traffic.

One solution to this situation is the unsourced random access (URA) scheme, which is first proposed in \cite{8006984}. In URA, all the users share a common codebook and choose codewords from the codebook as their messages based on a specific strategy, which can accommodate a large number of potential users. The BS only recovers the list of transmitted messages regardless of the corresponding users' IDs, thus leading to the so-called unsourced
property. There has been some related work in additive white Gaussian multiple access channel (GMAC) and fading channel\cite{8006985,8849802,amalladinne2018coded,9013278,9110507}. In \cite{8006985}, a low complexity coding scheme is proposed which requires a lower energy-per-bit than some traditional schemes such as ALOHA and treating interference
as noise (TIN). A cascaded system is proposed in \cite{8849802}, which includes inner and outer codes. To reduce the complexity, the outer tree encoder \cite{amalladinne2018coded} first splits the data into several slots with a large number of parity bits added. Within each slot, the messages are encoded by the sparse regression
codes (SPARCs). The inner compressed sensing (CS) based AMP decoder recovers the slot-wise transmitted messages. The outer tree decoder then stitches the decoded messages across different slots according to the prearranged parity. The complexity of the tree decoder increases exponentially with the number of slots. In \cite{9013278}, the user's data is divided into two parts. The former part acts as a preamble sequence for recovering the number of active users as well as the corresponding interleaving patterns. The latter part is a low-density-parity-check (LDPC) coded interleave-division multiple access (IDMA) scheme. In \cite{9110507}, a joint fading coefficient estimation and LDPC decoding scheme is proposed based on BP.

Besides the above works, the study of massive URA in multiple-input multiple-output (MIMO) system has also drawn increasing attention. The BS equipped with multiple antennas adds extra dimensions for the received signal, thus contributing to the signal detection and the access of a massive number of users.
In \cite{8437359,9049039}, a covariance-based maximum likelihood (CB-ML) scheme is proposed in MIMO massive URA system. The CB-ML scheme also adapts a concatenated coding scheme with the aforementioned tree code as the outer code, of which the inner decoder adapts a non-Bayesian method (i.e., ML detection) based on the covariance of the received signal. The covariance-based property of CB-ML makes it unable to acquire channel information and thus unable to combine with LDPC. The performance of the above CB-ML scheme degrades dramatically when the number of antennas is less than that of the active users. Besides, the CB-ML scheme has low spectral efficiency and high complexity because of the tree code.
In \cite{shyianov2020massive}, users' messages are split into several slots without adding parity bits. By clustering together the slot-wise recovered channels, the decoded slot-wise messages can be stitched correspondingly. However, the users' channels in \cite{shyianov2020massive} are assumed to remain constant over all the slots (i.e., $6798$ channel uses in their simulation). This assumption is difficult to hold in practice. In this paper, we propose an improved scheme with enhanced spectral efficiency and performance as well as low complexity for MIMO massive URA.

\subsection{Contributions}
Motivated by all the aforementioned works, we propose a joint SPARC-LDPC coding scheme for MIMO massive URA, which can accommodate a large number of active users with a reasonable antenna array size at the BS. In our scheme, we first split the information into two parts. The former part is encoded by means of SPARC to pick codewords from a fixed codebook. The AMP decoder recovers information in the former part and the corresponding channel coefficients as well as interleaving patterns. The latter part is encoded by LDPC codes and then interleaved by the IDMA scheme. With the channel coefficients and interleaving patterns estimated by the former part, information in the latter part can be obtained based on BP joint with successive interference cancellation (SIC).

Compared with the state-of-the-art CB-ML scheme investigated in \cite{8437359,9049039}, our scheme has better performance when the number of antennas at the BS is smaller than that of active users. The complexity of our scheme is with the order $\mathcal{O}\left(2^{B_p}ML+\widehat{K}ML\right)$ and lower than the CB-ML scheme. Moreover, our scheme has higher spectral efficiency (i.e., $15$ times larger) as we only split messages into two parts.

\section{System Model}
Consider the uplink of a single-cell cellular network consisting of $K_{tot}$ potential users, in which ${K_a}$ users are active in a slot. The BS is equipped with $M$ antennas. Each user is equipped with a single antenna and has $B$ bits of information to be coded and transmitted in a block-fading channel. As mentioned above, all the users share a common codebook in URA, denoted by $\mathbf{A} \in {\mathbb{C}^{L \times {2^B}}}$. The power of each codeword $\mathbf{a}_k$ is constraint to $L$, i.e., $\left\| {{\mathbf{a}_k}} \right\|_2^2 = L$. Let ${i_k} \in \left[ {1:{2^B}} \right]$ denote the  message indice of user $k$. Then $\mathbf{a}_{i_k}$ is the coded message of user $k$. In this work, the map between messages and codewords is a SPARC scheme. The main idea of SPARC is to map the information to a sparse vector and then choose the corresponding codeword from a fixed codebook according the sparse vector. The corresponding received signal can be written as
\begin{equation}
\mathbf{Y} = \sum\limits_{k \in {K_a}} {{\mathbf{a}_{{i_k}}}{\mathbf{h}_{{i_k}}^T}}  + \mathbf{Z} = \mathbf{A}\mathbf{\Phi} \mathbf{H} + \mathbf{Z}
\label{equ-1}
\end{equation}
where $\mathbf{\Phi}  \in {\left\{ {0,1} \right\}^{{2^B} \times {2^B}}}$ denotes the diagonal binary selection matrix. The positions of one in $\mathbf{\Phi}$ indicate that there are users whose messages are mapped there. However, the users' IDs are unknown to the BS, thus leading to the so-called unsourced property. Notice $\mathbf{\Phi}$ is a sparse matrix which is a consequence of the SPARC and can be recovered by CS techniques. $\mathbf{H} \in {\mathbb{C}^{{2^B} \times M}}$ denotes the MIMO channel coefficient matrix. $\mathbf{Z} \in {\mathbb{C}^{{L} \times M}}$ denotes additive white Gaussian noise (AWGN) and is distributed as $\mathcal{CN}\left( {0,{\sigma _n^2}\mathbf{I}} \right)$.

Let $\mathcal{L}$ denote the set of recovered messages at the BS. The performance in URA is evaluated by the probability of misdetection and false alarm, denoted by $p_{md}$ and $p_{fa}$ respectively, which are given by:
\begin{align}
\label{equ-2}p_{md} &= \frac{1}{{{K_a}}}\sum\limits_{k \in {\mathcal{K}_a}} {P\left( {{\mathbf{a}_{{i_k}}} \notin \mathcal{L}} \right)} \\
\label{equ-3}f_{fa} &= \frac{{\left| {\mathcal{L}\backslash \left\{ {{\mathbf{a}_{{i_k}}}:k \in {\mathcal{K}_a}} \right\}} \right|}}{{\left| \mathcal{L} \right|}}
\end{align}
where $\left|  \cdot  \right|$ denotes the Hamming weight. In this system, the code rate $R=\frac{B}{L}$ and the spectral efficiency $\mu =\frac{{B \cdot {K_a}}}{{L \cdot M}}$.

\section{Proposed Scheme}
Notice the dimension of the codebook $\mathbf{A}$ increases exponentially with the information length $B$. To reduce the complexity, a tree code \cite{amalladinne2018coded} is implied to split the information into several slots in the CB-ML scheme\cite{9049039}. As mentioned above, the tree code greatly reduces the code rate and spectral efficiency as a large number of parity bits are added during the encoding process. Besides, the complexity of the tree decoding process increases exponentially with the number of slots. In what follows below, we propose a low-complexity joint SPARC-LDPC coding scheme which keeps the spectral efficiency in a relatively high level.

\subsection{Encoder}
Motivated by \cite{9013278}, we divide the $B$ bits message into two parts, where the former part is coded with SPARC and tasked to recover part of the messages, the number of active users, the interleaving patterns and most importantly, the MIMO channel based on CS. The latter part is coded with LDPC codes and can be decoded by BP with the channel estimated by the former part. For clarity, we denote the former and latter parts as CS and LDPC parts, respectively. The total $B$ bits are split into two parts of $B_p$ and $B_c$, respectively, where ${B_p} \ll {B_c}$. Correspondingly, the total $L$ channel uses are split into $L_p$ and $L_c$, respectively, where ${B_p}+{B_c}=B$ and $L_p+L_c=L$. Compared with the existing scheme of \cite{9049039}, to support the same amount of users, the required number of channel uses $L$ in our scheme is much shorter, resulting in higher spectral efficiency. An example is shown in Section \ref{sec-IV}.

For the CS encoding part, we use SPARCs to construct a sparse selection matrix $\Phi$ as shown in \eqref{equ-1}.  Each user chooses a codeword from the codebook $\mathbf{A} = \left[ {{\mathbf{a}_1},{\mathbf{a}_2},...,{\mathbf{a}_{{2^{{B_P}}}}}} \right]\in {\mathbb{C}^{{L_p} \times {2^{B_p}}}}$. Let ${i_k} \in \left[ {1:{2^{B_p}}} \right]$ denote the message indice of user $k$. Then the $B_p$ bits of user $k$ are coded to the codeword $\mathbf{a}_{i_k}$ with power constraint $\left\| {{\mathbf{a}_{i_k}}} \right\|_2^2 = L_p$.  A user's message indice $i_k$ and the $B_p$ bits of information is a one-to-one mapping. If $i_k$ is recovered then the corresponding $B_p$ bits of information can be recovered automatically. Users who has the same $B_p$ bits of information will choose the same codewords, in which case their corresponding channel coefficients cannot be estimated properly. This collision has been considered in our numerical results.

\begin{figure}[htbp]
	\vspace{-0.2cm}
	\centerline{\includegraphics[width=0.5\textwidth]{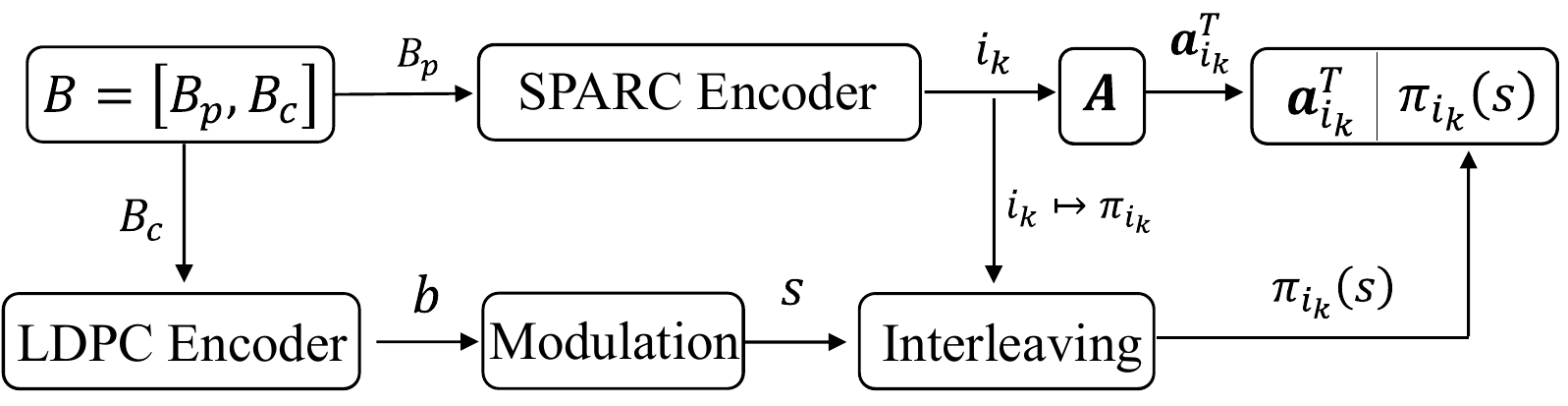}}
	\caption{The proposed encoding scheme.}
	\label{pic-1}
	\vspace{-0.2cm}
\end{figure}

For the LDPC encoding part, user $k$'s $B_c$ bits of information are encoded to an $\left( {{L_c},{B_c}} \right)$ LDPC code  $\mathbf{b} \in {\left\{ {1, 0} \right\}^{{L_c} \times 1}}$ and then modulated by binary phase shift keying (BPSK), denoted by $\mathbf{s} \in {\left\{ {1, - 1} \right\}^{{L_c} \times 1}}$. Then $\mathbf{s}$ is interleaved by a random interleaver ${\pi _{{i_k}}}$ with interleaving pattern $i_k$. This is nearly an IDMA scheme because the interleaving patterns for most of users are different as the message indices $i_k$ are different. The above encoding scheme is illustrated in Fig. \ref{pic-1}. After the two parts of encoding, the final coded message of user $k$ is $\mathbf{v} = {\left[ {\mathbf{a}_{{i_k}}^T,{\pi _{{i_k}}^T}\left( \mathbf{s} \right)} \right]}^T$. The received signal in this coding scheme is $\mathbf{Y} = \mathbf{V}\mathbf{H}+\mathbf{Z}$, where $\mathbf{V} = \left[ {{\mathbf{v}_1},{\mathbf{v}_2},...,{\mathbf{v}_{{K_a}}}} \right] \in {\mathbb{C}^{L \times {K_a}}}$ denotes users' coded messages.

\subsection{Decoder}
 The overall decoder consists of CS and LDPC decoding parts, which is summarized in Algorithm \ref{alo-1}. The details of the decoding process are given below.

\subsubsection{CS decoding part}
The CS decoder recovers the user's message indice $i_k$ and the corresponding channel coefficient vector $\mathbf{h}_{i_k}$. Then the $B_p$ bits of information and interleaving pattern ${\pi _{{i_k}}}$ can be obtained by $i_k$. Rewrite the received signal in (\ref{equ-1})
\begin{equation}
\mathbf{Y}_p ={\mathbf{Y}_{1:{L_p},:}}= \sum\limits_{k \in {\mathcal{K}_a}} {{\mathbf{a}_{{i_k}}}{\mathbf{h}_{{i_k}}^T}}  + \mathbf{Z}_{1:{L_p},:} = \mathbf{A}\mathbf{X} + {\mathbf{Z}_{1:{L_p},:}}
\label{equ-5}
\end{equation}
where $\mathbf{Y}_p \in {C^{{L_P} \times M}}$ is the first $L_p$ rows of received signal $\mathbf{Y}$ and $\mathbf{X}=\mathbf{\Phi} \mathbf{H} \in {\mathbb{C}^{{2^{{B_P} \times M}}}}$ is a row sparse matrix which can be recovered by AMP as follows.
\begin{align}
{\mathbf{X}^{t + 1}} &= {\eta _t}\left( {{\mathbf{A}^H}{\mathbf{Z}^t} + {\mathbf{X}^t}} \right) \label{equ-6}\\
{\mathbf{Z}^{t + 1}} &= \mathbf{Y}_p - \mathbf{A}{\mathbf{X}^{t + 1}} + \frac{{{2^{B_p}}}}{L}{\mathbf{Z}^t}\left\langle {{\eta _t}^\prime \left( {{\mathbf{A}^H}{\mathbf{Z}^t} + {\mathbf{X}^t}} \right)} \right\rangle \label{equ-7}
\end{align}
where ${\eta _t}\left(  \cdot  \right):{\mathbb{C}^{^{{B_p} \times M}}} \to {\mathbb{C}^{^{{B_p} \times M}}}$ is the denoiser which is a row-wise function. ${\mathbf{Z}^{t + 1}}$ is the corresponding residual at the $\left({t+1}\right)$-th iteration. The last term of ${\mathbf{Z}^{t + 1}}$ is called Onsager term which is to adjust the correlation problem during each iteration and the involved ${\eta _t}^\prime \left(  \cdot  \right)$ is the first-order derivative of ${\eta _t}\left(  \cdot  \right)$. For details, please refer to \cite{8323218} and \cite{donoho2009message}. Note the sparsity (i.e., ${{{K_a}} \mathord{\left/
		{\vphantom {{{K_a}} {{2^{B_p}}}}} \right.\kern-\nulldelimiterspace} {{2^{B_p}}}}$) is needed in AMP\cite{8323218} as a prior information which may be unknown in real scenarios. Nevertheless, we find that AMP algorithm is insensitive to the sparsity. So for fair comparison with the CB-ML scheme, we use $0.1$ as the input sparsity no matter what the actual sparsity is.

The CS decoding process ends when the mean square error (MSE) of $\left({{\mathbf{X}^{t + 1}}-{\mathbf{X}^{t}}}\right)$ is small enough (i.e., lower than a certain threshold) or the maximum number of iterations is reached. Finally, the CS decoder outputs the set of estimated message indices $\mathcal{L}= \left\{ {{i_1},{i_2},...{i_K}} \right\}$, the set of the corresponding subscripts, $\mathcal{K}$, as well as the corresponding channel vector $\left\{ {{{\widehat {\mathbf{X}}}_{{i_{\rm{k}}},:}},k \in \mathcal{K}} \right\}$, and we denote it by $\widehat {\mathbf{H}} \in {\mathbb{C}^{K \times M}}$ which will be used for the LDPC decoding part.

\subsubsection{LDPC decoding part}
\begin{figure*}[ht]
	\centerline{\includegraphics[width=0.8\textwidth]{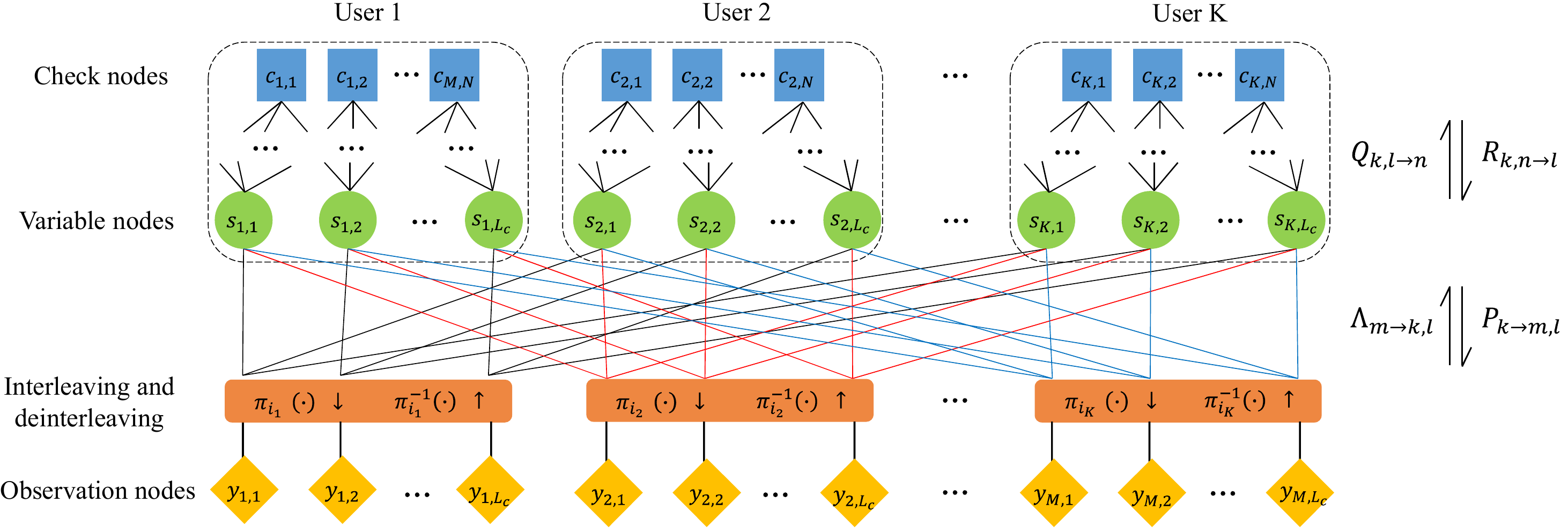}}
	\caption{Factor graph for LDPC decoding.}
	\label{pic-2}
	\vspace{-0.5cm}
\end{figure*}

The received signal of the LDPC part can be written as
\begin{equation}
{\mathbf{Y}_c} = {\mathbf{Y}_{{L_p} + 1:{L_c},:}} = \sum\limits_{k \in {\mathcal{K}_a}} {{\pi _{{i_k}}}\left( {{\mathbf{s}_k}} \right){\mathbf{h}_{{i_k}}^T} + {\mathbf{Z}_{_{{L_p} + 1:{L_c},:}}}}
\label{equ-8}
\end{equation}
where $\mathbf{Y}_c \in {C^{{L_c} \times M}}$ is the last $L_c$ rows of $\mathbf{Y}$. The LDPC decoder is tasked to recover the last $B_c$ bits of information based on $\mathbf{Y_c}$ and $\widehat{\mathbf{H}}$ using the low-complexity iterative BP algorithm. The BP based decoding scheme can be illustrated by a factor graph as shown in Fig. \ref{pic-2}. The subscript $N$ in Fig. \ref{pic-2} denotes the number of check nodes in the LDPC code, which corresponds to the number of rows of the LDPC check matrix. Other subscripts are consistent with the aforementioned. Three types of nodes are shown in the factor graph. The check nodes (blue color) and variable nodes (green color) as well as the edges connecting them constitute to the Tanner graph in LDPC fields. The observation nodes (yellow color) correspond to the elements of received signal $\mathbf{Y}_c$.

There are edges in the factor graph which represent the connections between nodes.  The edges between check nodes and variable nodes are determined by the LDPC check matrix which cannot be clearly marked in the graph. The edges between variable nodes and observation nodes are simply determined by \eqref{equ-8} though looking complicated. For example, the observation node $y_{1,1}$ is connected to the first variable nodes of all users (i.e., $s_{k,1},k=1,2,...K$). Correspondingly, the variable node $s_{1,1}$ is connected to the first observation nodes from all antennas (i.e., $y_{m,1},m=1,2,...,M$). In BP algorithm, messages are passed along these edges. The types of messages are listed below.
\begin{itemize}
\item ${R_{k,n \to l}}$: Messages from check node $c_{k,n}$ to variable node $s_{k,l}$.
\item ${Q_{k,l \to n}}$: Messages from variable node $s_{k,l}$ to check node $c_{k,n}$.
\item ${P_{k \to m,l}}$: Messages from variable node $s_{k,l}$ to observation node $y_{m,l}$.
\item ${\Lambda _{m \to k,l}}$: Messages from observation node $y_{m,l}$ to variable node $s_{k,l}$ .
\end{itemize}

Note that interleaving and deinterleaving are needed for ${P_{k \to m,l}}$ and ${\Lambda _{m \to k,l}}$, respectively, of which the patterns can be obtained from the CS decoder. The interleaver is the connection between SPARC and LDPC. If the LDPC decoder uses the interleaver $\pi_{i_k}$ and successfully decodes the $B_c$ bits of message, the corresponding former $B_p$ bits of message can be obtained by mapping from $i_k$. Here we give details of the updating scheme of the messages for BPSK modulated system, which are listed above.
 \begin{align}
 {\Lambda _{m \to k,l}} &= \log \frac{{P\left( {{y_{m,l}}|\mathbf{H},{s_{k,l}} = 1} \right)}}{{P\left( {{y_{m,l}}|\mathbf{H},{s_{k,l}} =  - 1} \right)}} \\\notag
 						&= \frac{2}{{\sigma _{{z_{k,l}}}^2}}\mathcal{R}\left( {h_{m,k}^ * \left( {{y_{m,l}} - {\mu _{{z_{k,l}}}}} \right)} \right)
 \label{equ-9}
 \end{align}
where $\mathbf{H}$ is the channel matrix estimated in the CS decoding part and $\mathcal{R}\left(  \cdot  \right)$ denotes the real part of a complex number. In fact,  $ {\Lambda _{m \to k,l}}$ is the log-likelihood ratio (LLR) of variable node $s_{k,l}$ observed at node $y_{m,l}$. This is a TIN scheme as the received signal $y_{m,l}$ is given by
\begin{align}
	{y_{m,l}} &= \sum\limits_{j \in \mathcal{K}} {{h_{m,j}}{s_{j,l}} + {n_{m,l}}}\\\notag
	  		  &={h_{m,k}}{s_{k,l}} + \sum\limits_{j \in \mathcal{K} \backslash  k} {{h_{m,j}}{s_{j,l}} + {n_{m,l}}} ={h_{m,k}}{s_{k,l}}+ z_{k,l}
\end{align}

where $n_{m,l}$ is the Gaussian noise with zero mean and variance ${\sigma _n^2}$, and $\mathcal{K} \backslash k$ denotes all the numbers in range $\left[1:K\right]$ except $k$. In what follows, we use $\left[M\right]$ to denote the set of integers $\{1,2,...,M\}$. The Gaussian noise $n_{m,l}$ and the interference from other users are all treated as noise denoted by $z_{k,l}$, which is a Gaussian variable with mean ${\mu _{{z_{k,l}}}}$ and variance ${\sigma _{{z_{k,l}}}^2}$ given below.
\begin{align}
	{\mu _{{z_{k,l}}}}    &= \sum\limits_{j \in \mathcal{K} \backslash k} {{h_{m,j}}\left( {2{P_{j \to m,l}} - 1} \right)} \\
	\sigma _{{z_{k,l}}}^2 &= 4\sum\limits_{j \in \mathcal{K} \backslash k} {{{\left| {{h_{m,j}}} \right|}^2}\left( {1 - {P_{j \to m,l}}} \right){P_{j \to m,l}}}+\sigma_n^2
\end{align}
where ${P_{j \to m,l}}$ denotes the probability of $s_{j,l}=1$, and is initialized to $0.5$. The update of ${P_{j \to m,l}}$ is given by
\begin{equation}
	{P_{k \to m,l}} = \frac{{\exp \left( {\sum\limits_{j \in M \backslash m} {{\Lambda _{j \to k,l}}}  + \sum\limits_{j \in {\mathcal{N}_c}\left( {k,l} \right)} {{R_{k,j \to l}}} } \right)}}{{1 + \exp \left( {\sum\limits_{j \in M \backslash m} {{\Lambda _{j \to k,l}}}  + \sum\limits_{j \in {\mathcal{N}_c}\left( {k,l} \right)} {{R_{k,j \to l}}} } \right)}}
\end{equation}

The updates of other two messages in LDPC decoding process are given by
\begin{align}
	{Q_{k,l \to n}} &= \sum\limits_{j \in M} {{\Lambda _{j \to k,l}} + } \sum\limits_{j \in {\mathcal{N}_c}\left( {k,l} \right) \backslash n} {{R_{k,j \to l}}} \\
	{R_{k,n \to l}} &= 2{\tanh ^{ - 1}}\left( {\prod\limits_{j \in {\mathcal{N}_v}\left( {k,n} \right) \backslash l} {\tanh \left( {\frac{{{Q_{k,j \to n}}}}{2}} \right)} } \right)	
\end{align}
where ${\mathcal{N}_c}\left( {k,l} \right) \backslash n$ denotes the set of check nodes connected to $s_{k,l}$ except $c_{k,n}$, and ${\mathcal{N}_v}\left( {k,n} \right) \backslash l$ denotes the set of variable nodes connected to $c_{k,n}$ except $s_{k,l}$. ${R_{k,n \to l}}$ is initialized to $0$. The LLR of the variable node $s_{k,l}$ at the end of an iteration is given by
\begin{equation}
	{L_{k,l}} = \sum\limits_{j \in M} {{\Lambda _{j \to k,l}} + \sum\limits_{j \in {\mathcal{N}_c}\left( {k,l} \right)} {{R_{k,j \to l}}} }.
\end{equation}

The decoded bit ${\hat s_{k,l}}$ is $1$ when ${L_{k,l}}>0$ and $0$ otherwise. Let $ \mathbf{S} \in {\left\{ {0,1} \right\}^{{L_c} \times K}}$ and $ \mathbf{H}_c \in {\left\{ {0,1} \right\}^{{N} \times L_c}}$  denote the decoded messages and LDPC check matrix respectively. The LDPC decoding process ends when $mod \left({\mathbf{H}_c \mathbf{S},2}\right)=\mathbf{0}$ or the maximum number of iterations is reached. Note the estimated number of active users $K$ is not guaranteed to be equal to $K_a$. Therefore, not all the decoded messages in $\mathbf{S}$ satisfy the check. Let $\widehat{\mathcal{S}}=\left\{\mathbf{s}_k, k \in \widehat{\mathcal{K}}\right\}$ denotes the set of the successfully decoded messages and  $\widehat{\mathcal{K}}$ is the set of the corresponding subscripts. Obviously, $\left| {\widehat {\mathcal{K}}} \right| \le {K_a} \le K$. Hence the LDPC part can reduce the probability of false alarm. Finally, the LDPC decoder outputs $\widehat{\mathcal{S}}$ and  $\widehat{\mathcal{K}}$. To further improve the performance, we combine the LDPC decoder with SIC and we denote it by LDPC-SIC. LDPC-SIC works as follows.

Let $\widehat {\mathbf{H}} \in {C^{K \times M}}$ denote the channel matrix estimated by the CS decoder and $\mathcal{K}$ is the set of the corresponding subscripts. Let ${\hat {\mathcal{S}}_0} = \emptyset $ and ${\hat {\mathcal{K}_0}} = \emptyset $ respectively denote the sets of decoded messages and the corresponding subscripts obtained by the LDPC decoder, which are initialized to empty sets. With $\mathbf{Y}_c$ and $\widehat {\mathbf{H}} $, the LDPC decoder outputs the set of decoded messages  ${\widehat {\mathcal{S}}} $ and the corresponding subscripts ${\widehat {\mathcal{K}}}$. Then we have ${\hat {\mathcal{S}}_0} \leftarrow {\hat {\mathcal{S}}_0} \cup \widehat {\mathcal{S}}$, ${\hat {\mathcal{K}}_0} \leftarrow {\hat {\mathcal{K}}_0} \cup \widehat {\mathcal{K}}$ and ${\mathbf{H}} = {\hat {\mathbf{H}}_{k,:}}$ for $k \in {\mathcal{K}} \backslash {\hat {\mathcal{K}}_0}$. The residual signal is updated by
\begin{equation}
	\mathbf{Y} = {\mathbf{Y}_c} - \sum\limits_{k \in {{\hat {\mathcal{K}}}_0}} {\left( {2{\mathbf{s}_k} - 1} \right)\mathbf{h}_k}
	\label{equ-17}
\end{equation}
where $\mathbf{s}_k \in {\left\{ {0,1} \right\}^{{L_c} \times 1}}$ is the $k$-th codeword in ${\hat {\mathcal{S}}_0}$. Note $\mathbf{s}_k$ needs to be BPSK modulated when calculating $\mathbf{Y}$. $\mathbf{h}_k$ is the $k$-th row vector in $\widehat {\mathbf{H}}$.  $\mathbf{Y}$ and $\mathbf{H}$ are inputs to the LDPC decoder for next decoding. This iterative process ends when ${\hat {\mathcal{K}}} = \emptyset $ or ${\mathcal{K}} \backslash {\hat {\mathcal{K}}_0} =\emptyset$. The overall decoding scheme is shown in Algorithm \ref{alo-1}.
\begin{algorithm}
\caption{Joint CS and LDPC-SIC decoder for URA}
\label{alo-1}
\begin{algorithmic}
\STATE{{\bf Input}: $\mathbf{Y}_p$,  $\mathbf{A}$}
\STATE{$ \triangleright $ CS decoder}
\STATE{Output: $\widehat {\mathbf{H}}$,  $\mathcal{L}= \left\{ {{i_k},k \in \mathcal{K}} \right\}$}	
\STATE{Initialize: $\mathbf{Y}=\mathbf{Y}_c$,~~$\mathbf{H} =\widehat {\mathbf{H}}$,~~${\widehat {\mathcal{S}}_0} \leftarrow \emptyset $,~~${\widehat {\mathcal{K}}_0} \leftarrow \emptyset $}
\REPEAT
\STATE{Input: $\mathbf{Y}$, $\mathbf{H}$, noise var ${\sigma _n^2}$}
\STATE{Initialize: ${P_{k \to m,l}}=0.5$, ${R_{k,n \to l}}=0$ for all $k \in \left[K\right]$, $m \in \left[M\right]$,  $l \in \left[L_c\right]$ and $n \in \left[N\right]$.}
\STATE{$ \triangleright $ LDPC decoder}
\STATE{Output: ${\widehat {\mathcal{S}}} $,  ${\widehat {\mathcal{K}}}$}
\STATE{${\widehat {\mathcal{S}}_0} \leftarrow {\widehat {\mathcal{S}}_0} \cup \widehat {\mathcal{S}}$, ${\widehat {\mathcal{K}}_0} \leftarrow {\widehat {\mathcal{K}}_0} \cup \widehat {\mathcal{K}}$}
\STATE{${\mathbf{H}} = {\widehat {\mathbf{H}}_{k,:}}$ for $k \in {\mathcal{K}} \backslash {\widehat {\mathcal{K}}_0}$}
\STATE{$	\mathbf{Y} = {\mathbf{Y}_c} - \sum\limits_{k \in {{\hat {\mathcal{K}}}_0}} {\left( {2{\mathbf{s}_k} - 1} \right)\widehat {\mathbf{H}}_{k,:}} $}
\UNTIL{${\widehat {\mathcal{K}}} = \emptyset $ or ${\mathcal{K}} \backslash {\widehat {\mathcal{K}}_0} =\emptyset$}
\STATE{Return: ${\widehat {\mathcal{S}}_0}$, ${\widehat {\mathcal{K}}_0}$}
\end{algorithmic}	
\end{algorithm}

After the LDPC decoder outputs the subscript set of the decoded messages ${\widehat {\mathcal{K}}_0}$, the stitching of two parts of messages is easy. Let $\widehat{\mathcal{L}}=\left\{ {{i_k},k \in {{\widehat{\mathcal{K}} }_0}} \right\}$ denotes the message indices corresponding to users in $\widehat{\mathcal{K}}_0$.  $\widehat{\mathcal{L}}$ can be directly mapped to the $B_p$ bits of information.  Let $\widehat{\mathcal{S}}$ denotes the set of $B_c$ bits of information by removing the redundant LDPC check bits of ${\widehat {\mathcal{S}}_0}$. Then these two parts of information with the same subscript $k \in {\widehat {\mathcal{K}}_0}$ can be stitched together. Then the probability of misdetection and false alarm can be obtained by \eqref{equ-2} and \eqref{equ-3}.

\section{Numerical Results}\label{sec-IV}
We compare by numerical results the performance of the proposed LDPC and LDPC-SIC schemes with the CB-ML scheme of \cite{9049039} in various signal noise ratio (SNR) regions and with different number of antennas. The LDPC scheme refers to the LDPC decoder without SIC. The parameter settings in our simulation are shown in TABLE \ref{tab1}. In the LDPC scheme, $\left({L_c,B_c}\right)$ means the $B_c$ bits of information are coded to $L_c$ bits of LDPC code. The LDPC and LDPC-SIC decoders share the same parameter settings. Both are the $\left(3,6\right)$-regular LDPC code and the code rate is $0.5$. In the tree code scheme, the $96$ bits of information are split into $S=24$ slots with slot length $J=16$ bits. The first slot has $16$ bits of information and no parity bit. The following second to $21$th slots have $4$ bits of information and $12$ bits of parity, and the final $3$ slots have $16$ bits of parity and no information bit.  For tree encoding and decoding details, please refer to \cite{amalladinne2018coded}.

 \begin{table}[h]{}
 	\renewcommand{\arraystretch}{1.0}
 	\vspace{-0.2cm}
		\caption{Parameter Settings}
	\begin{center}
		\vspace{-0.2cm}	
		\begin{tabular}{|c|c|c|}
			\hline	
			& CB-ML & LDPC, LDPC-SIC\\
			\hline
			${E_b}\slash{N_0}$ (dB) & \multicolumn{2}{c|}{$\left[{20:30}\right]$} \\
			\hline
			$M$ & \multicolumn{2}{c|}{$\left[ {50:100} \right]$} \\
			\hline
			$K_a$ & \multicolumn{2}{c|}{$100$} \\
			\hline
			$L_p$ & \multicolumn{2}{c|}{$300$} \\
			\hline
			$L_c$ &{$\backslash$} & $200$\\
			\hline
			$B$ & $96$ & \multicolumn{1}{c|}{$96$,~~$B_p=16$,~~$B_c=80$} \\
			\hline
			LDPC & {$\backslash$} &  $\left({L_c,B_c}\right)$\\
			\hline
			tree code & {\tabincell{c}{$S=24$, $J=16$}}& {$\backslash$}\\
			\hline
			channel uses & $ S \times {L_p}=7200$ & $L_p+L_c=500$\\
			\hline
			code rate & $0.013$ &  $0.1920$ \\
			\hline			
		\end{tabular}
		\label{tab1}
		\vspace{-0.4cm}
	\end{center}
\end{table}

As is mentioned above, our scheme requires a much smaller number of channel uses compared with the CB-ML scheme of \cite{9049039}.  As is shown in TABLE \ref{tab1}, the CB-ML scheme needs $7200$ channel uses while our LDPC and LDPC-SIC schemes only need $500$ channel uses. As a consequence, the code rate of the proposed LDPC scheme is nearly $15$ times lager than that of the CB-ML scheme. Let $R$ denote the code rate. The energy per symbol $E_s$ of the coded messages is $E_s = {{E_b}\slash R}$. This is a fair comparison because both the transmitted data $B$ and the energy per bit $E_b$ are the same for CB-ML and our scheme. The large scale fading coefficient (LSFC) is set to $1$ and is known to all the above schemes and the noise variance ${N_0} \equiv 1$. The empirical threshold for all algorithms except AMP is set to 0.5. AMP has its own activity detection threshold according to formula (42) in \cite{6240094}.

\begin{figure}[h]
 	\vspace{-0.2cm}
	\centerline{\includegraphics[width=0.35\textwidth]{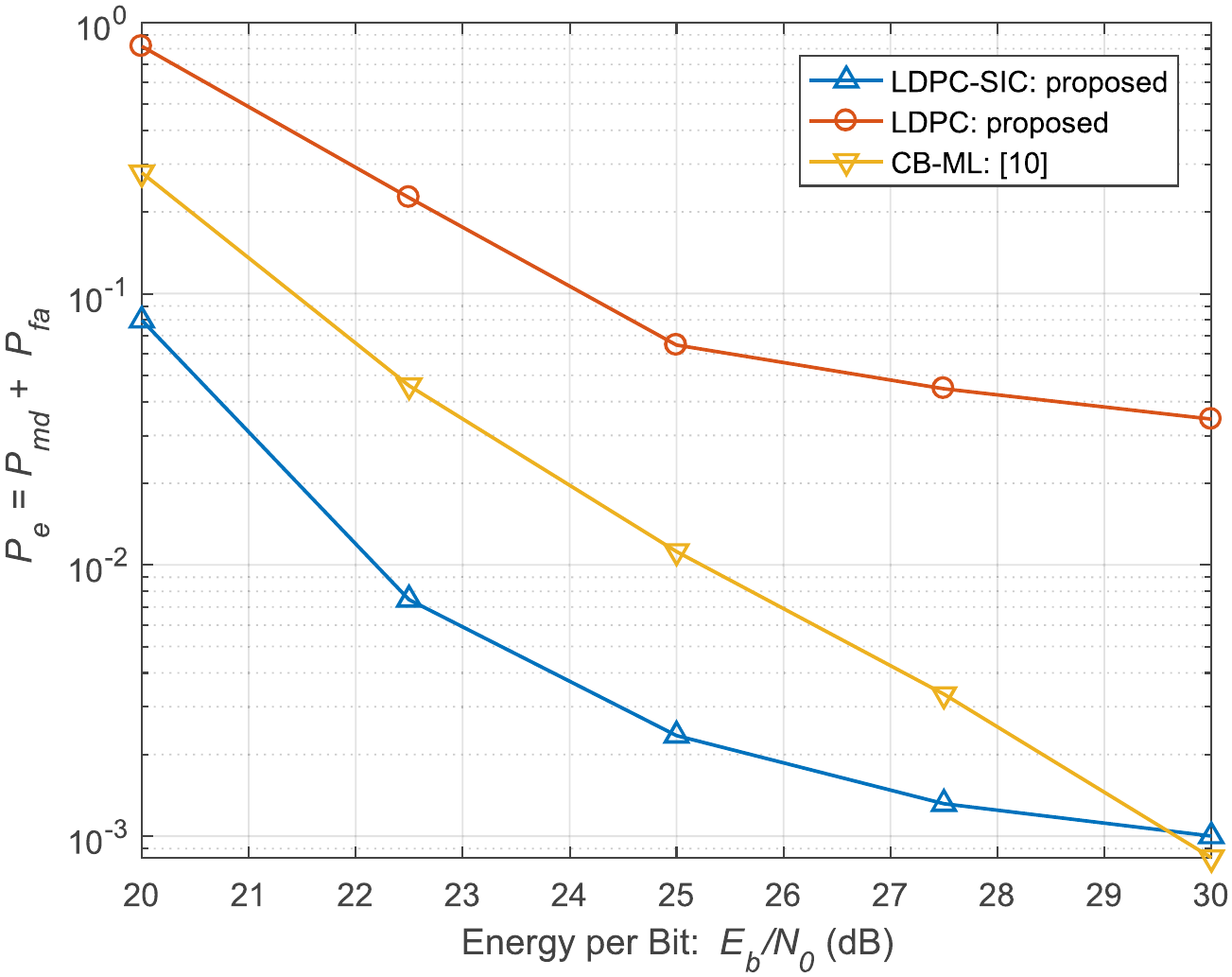}}
	\caption{Error probability $P_e=P_{md}+{P_{fa}}$ as a function of ${E_b}\slash{N_0}$, M=50. See TABLE \ref{tab1} for other parameter settings.}
	\label{pic-3}
	\vspace{-0.2cm}
\end{figure}

Fig. \ref{pic-3} shows how the error probability $P_e$ falls as a function of the energy per bit ${E_b}\slash{N_0}$ at $M=50$. In Fig. \ref{pic-3}, the proposed LDPC-SIC scheme outperforms the CB-ML scheme with a nearly $0.8$dB gap at ${E_b}\slash{N_0} = 22.5$ dB, while the LDPC scheme without SIC is a little bit worse. Besides, the spectral efficiency of LDPC-SIC is $0.384$ bps/Hz/RX which is nearly $15$ times larger than CB-ML, the spectral efficiency of which is $0.027$ bps/Hz/RX.

 \begin{table*}[ht]
	\renewcommand{\arraystretch}{1.2}
	\caption{Computational Complexity for LDPC, LDPC-SIC and CB-ML Schemes: Number of Floating Point Multiplications and Additions in Each Iteration}
	\vspace{-0.2cm}
	\begin{center}	
		\begin{tabular}{|c|c|c|c|}
			\hline
			\multicolumn{2}{|c|}{Schemes} & Floating Point Multiplications (FPM) & Floating Point Additions (FPA) \\
			\hline
			\multirow{2}{*}{CB-ML} & ML decoder& $2^J\left(3L^2+3L+4\right)+ML^2+L^2$ & $2^J\left(2L^2+1\right)+\left(M-1\right)L^2$\\
			\cline{2-4}
			& Tree decoder & \multicolumn{2}{c|}{$K\left( {\sum\nolimits_{j = 2}^S {{l_j}}  + \sum\nolimits_{j = 2}^{S - 1} {{l_{j + 1}}\sum\nolimits_{q = 2}^j {{K^{j - q}}\left( {K - 1} \right)\prod\nolimits_{i = q}^j {{2^{ - {l_i}}}} } } } \right)$}\\
			\hline
			\multirow{2}{*}{\tabincell{c}{LDPC\\LDPC-SIC}} &AMP decoder & $2\cdot2^{B_p}ML+ML+11\cdot2^{B_p}M+4M+2\cdot2^{B_p}+5$ & $2^{B_p}ML+4ML+6\cdot2^{B_p}M-2M-2^{B_p}$\\
			\cline{2-4}
			& LDPC decoder & $10\widehat{K}ML+2\widehat{K}NN_v+\widehat{K}M$ & $\left(13\widehat{K}-2\right)ML+\widehat{K}L\left(N_c-1\right)+2\widehat{K}MN_c$\\
			\hline
			\multicolumn{4}{l}{\tabincell{l}{Note: $K$ and $\widehat{K}$ are the number of active users estimated by the ML decoder and AMP decoder respectively. $N$ is the number of check bits in LDPC.\\ ~~~~~~~$N_c$ is the number of check nodes connected to a variable node and $N_v$ is the  number of variable nodes connected to a check node.}}
			
		\end{tabular}
		\label{tab2}
		\vspace{-0.4cm}
	\end{center}
\end{table*}

 \begin{figure}[h]
 	\vspace{-0.2cm}
	\centerline{\includegraphics[width=0.35\textwidth]{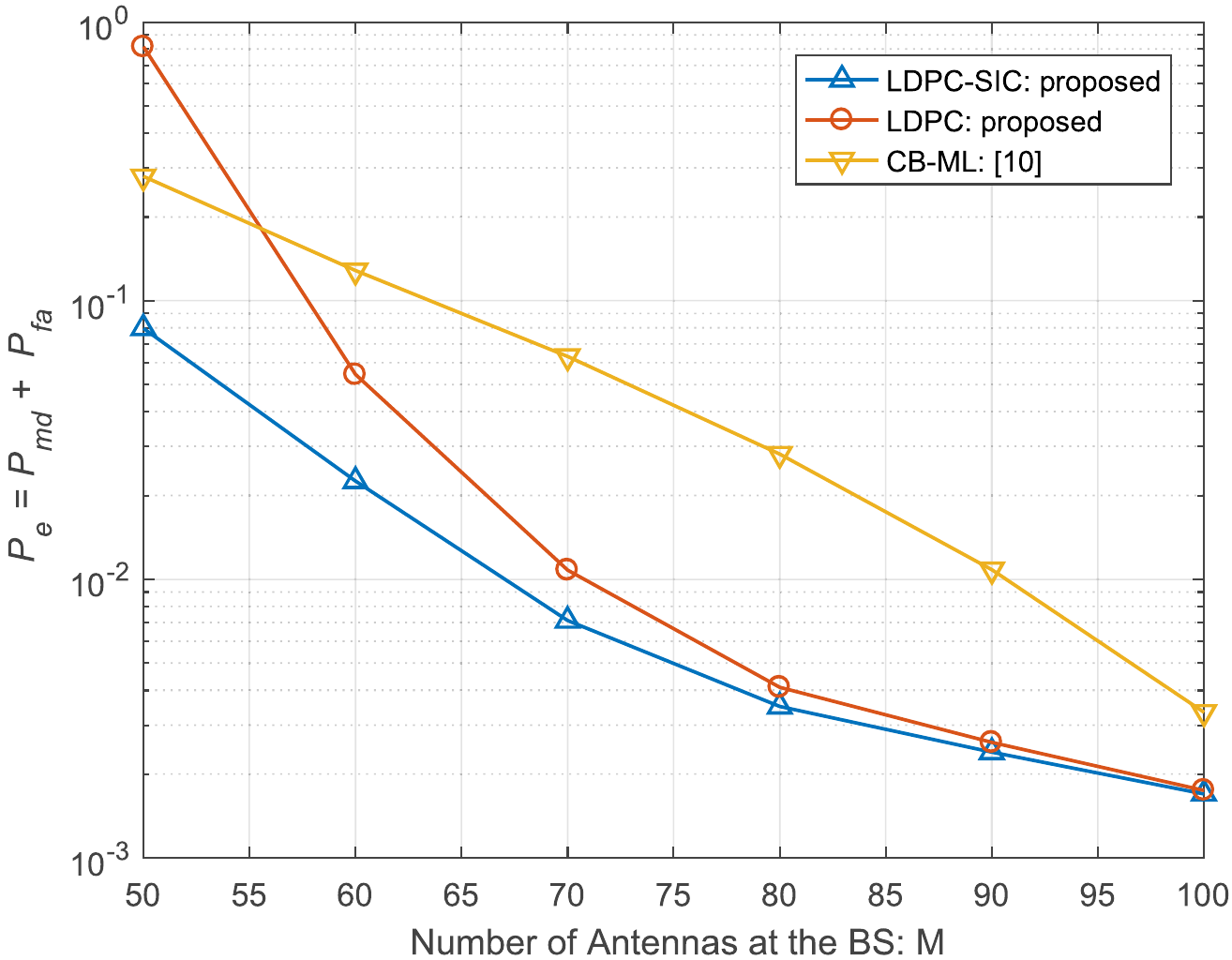}}
	\caption{Error probability $P_e=P_{md}+{P_{fa}}$ as a function of $M$, ${E_b}\slash{N_0}=20$ dB. See TABLE \ref{tab1} for other parameter settings.}
	\label{pic-4}
	\vspace{-0.2cm}
\end{figure}

Fig. \ref{pic-4} shows how the error probability $P_e$ falls as a function of the number of antennas $M$ at  ${E_b}\slash{N_0}=20$ dB. In Fig. \ref{pic-4}, the proposed LDPC and LDPC-SIC schemes both outperform CB-ML. There is a nearly $1$ dB gap between the LDPC-SIC and CB-ML scheme at $M=70$. Note that the SIC method contributes less as the number of antennas increases. As a consequence, the performance of LDPC and LDPC-SIC get closer. Besides, as is shown in Fig. \ref{pic-3} and \ref{pic-4}, our scheme doesn't perform that well and has a similar performance with the CB-ML in a high SNR region or large number of antennas. This is exactly because of the presence of the collision. But all in all, the above proposed schemes have higher spectral efficiency and outperform CB-ML in various values of transmitted power and number of antennas at the BS.

Moreover, the complexity analysis for above schemes is given in TABLE \ref{tab2}. The difference between the LDPC and LDPC-SIC schemes is that the latter needs more iterations, which is 3 to 4 times according to our simulation. The number of FPM and FPA of the ML decoder is with $\mathcal{O}\left(2^JL^2\right)$, of which the AMP decoder is with $\mathcal{O}\left(2^{B_p}ML\right)$. The complexity of the ML decoder is nearly the same order with that of the AMP decoder. But as a consequence of the coordinate descent, there are $2^J$ cycles in the ML decoding per iteration, which can only be computed successively. On the contrary, all computations in the AMP decoder can be performed in parallel. Hence, the AMP decoder has a lower time complexity than the ML decoder. Besides, the complexity of the tree decoder given by \cite{amalladinne2018coded} in the CB-ML scheme is with the order $\mathcal{O} \left(K^S\right)$, which increases exponentially with the number of slots, $S$. The complexity of the LDPC decoder in our scheme is linear with $\widehat{K}$, $M$, $L$ and with the order $\mathcal{O}\left(\widehat{K}ML\right)$. In conclusion, our scheme has complexity of order  $\mathcal{O}\left(2^{B_p}ML+\widehat{K}ML\right)$ and is lower than that of the CB-ML scheme.

\section{Conclusion}
In this paper, we propose a low-complexity SPARC-LDPC coding scheme for MIMO massive URA. Based on compressed sensing, belief propagation as well as successive interference cancellation, the proposed scheme outperforms the state-of-the-art CB-ML scheme when the number of active users is larger than that of the antennas at the BS. This is reasonable because the number of antennas at the BS is limited and less than the number of active users in a massive access scenario. The complexity of our scheme is with the order $\mathcal{O}\left(2^{B_p}ML+\widehat{K}ML\right)$ and lower than the CB-ML scheme. Furthermore, our scheme is able to recover users' information with a near $15$ times higher spectral efficiency than the CB-ML scheme. Our future work is to avoid collisions through some scheduling measures and take a message passing method between SPARC and LDPC modules.

\end{document}